\documentstyle[prl,aps]{revtex}
\begin{document}
\draft

\title{From Antiferromagnetism to Superconductivity:
Numerical Evidence for $SO(5)$ Symmetry
and Superspin Multiplets}

\author{Robert Eder, Werner Hanke}
\address{Institut f\"{u}r Theoretische Physik, Am Hubland,
D-97074 W\"{u}rzburg, Federal Republic of Germany}
\author{Shou-Cheng Zhang}
 \address{
Department of Physics, Stanford University, Stanford, CA 94305
}
 
\date{\today}

\maketitle
\begin{abstract}
In this work, we present numerical results which support $SO(5)$ symmetry as
a concept unifying superconductivity and antiferromagnetism in the
high-temperature superconductors. Using exact cluster diagonalization, we
verify that the low-energy states of the $t-J$ model, a widely used
microscopic model for the high $T_c$ cuperates, form $SO(5)$ symmetry
multiplets.
Our results show that the $d-$wave
superconducting ground states away from half-filling are obtained from the
higher spin states at half-filling through $SO(5)$ rotations.
\end{abstract}
\newpage
 
{\bf Introduction:}
Ever since the early days of quantum mechanics, group 
theoretical interpretation of spectroscopy revealed deep symmetry and 
profound unity of Nature. Atomic spectra can be fitted into irreducible 
representations (irreps) of $SO(3)$, and the regular patterns which emerged 
from this classification offered fundamental understanding of the periodic 
table. After the discovery of a large number of hadrons, the ``embarrassment 
of riches'' is removed by the classification of hadronic spectra into irreps 
of $SU(3)$ and this hidden regularity inspired the predictions of quarks, 
the fundamental building block of the universe. In our quest for 
understanding the fundamental design of Nature, the importance of symmetry 
can never be over-emphasized. 
 
In this article we report a different kind of spectroscopy, and its 
classification into a different kind of symmetry. The spectroscopy is 
performed on a computer, which numerically diagonalizes microscopic 
Hamiltonians, i. e. Hubbard and $t-J$ models widely 
believed\cite{anderson,zhangrice} to model 
high-$T_{c}$ superconductors. One of the universal features of high-$T_{c}$ 
superconductors is the close proximity between the antiferromagnetic (AF) 
and the $d-$ wave superconducting (dSC) 
phases\cite{scalapino,schrieffer,pines,dagotto,emery}. While an AF insulator 
appears to be diagonally opposite to a superconductor, their close proximity 
lead one of us (SCZ) to conjecture that they are in fact intimately related 
by an $SO(5)$ symmetry group which unifies them\cite{zhang}. In this theory, 
the AF 
and dSC order parameters are grouped into a single five component vector $% 
n_{a}$ called superspin. The transition from AF to dSC is viewed as 
a superspin flop transition as a function of the chemical potential or 
doping, where the direction of the superspin changes abruptly. 
 
While the $SO(5)$ symmetry was originally proposed in the context of an 
effective field-theory description of the high-$T_{c}$ superconductors, its 
predictions can actually be tested within microscopic models. First 
numerical evidence for the approximate $SO(5)$ symmetry of the Hubbard model 
came recently from exact diagonalizations of small-sized (10 sites) 
clusters\cite{hanke} 
studying dynamic correlations functions involving the AF/dSC 
rotation $\left( \pi \right) $ operators\cite{demler,zhang}. 
In this work we shall use the $t-J$ model because of its more limited
Hilbert space. Since the $t-J$ model explicitly projects out the
states in the upper Hubbard band, some of the 
questions\cite{greiter,reply,baskaran}
raised recently about the compatibility between the Mott Hubbard gap
and $SO(5)$ symmetry can also be answered explicitly. We use a 
general and direct recipe for checking microscopic Hamiltonians for $SO(5)$ 
symmetry, i.e. the concept of ``superspin multiplets". In particular, if 
there is an approximate $SO(5)$ symmetry of the microscopic model, the 
low-energy states of this model should fall into irreps of $SO(5)$. In a 
given quantum mechanical system, the direction of the $SO(5)$ superspin 
vector is quantized in a way similar to an ordinary $SO(3)$ spin, and the 
classically intuitive picture of the precession of the $SO(5)$ superspin 
vector under the influence of the chemical potential\cite{zhang,fradkin} 
can be 
identified with the equal level-spacing between the members of $SO(5)$ 
multiplets carrying different charge. Therefore, numerically identifying the 
low-lying states of the microscopic model with the $SO(5)$ irreps can lead 
to detailed understanding of the one-to-one correspondence and the level 
crossing between the excited states of the AF and the dSC states, and 
thereby lead us to the microscopic mechanism by which the AF changes into 
the dSC state. While finite size-calculations can not generally be used to
prove the existence of long range order in infinite systems, the spectroscopic
information about the $SO(5)$ symmetry can be used as input for the 
effective field theory\cite{zhang,burgess,arovas} which captures the low 
energy and long distance physics of the problem.
 
{\bf SO(5) Superspin Multipets, a Pyramid of Diamonds:} 
The $SO(5)$ Lie algebra is generated by ten 
operators $L_{ab}$ with $a,b=1,..,5$ and $a<b$. They obey the following 
commutation relation:  
\begin{eqnarray} 
\left[ L_{ab}, L_{cd} \right] = i( \delta_{ac} L_{bd} + \delta_{bd} L_{ac} - 
\delta_{ad} L_{bc} -\delta_{bc} L_{ad}) 
\end{eqnarray} 
$SO(5)$ is a rank two algebra, we can therefore choose total the
charge $Q\equiv 
L_{15}=\frac{1}{2}(N_e-M)$ and $z$ component of the total spin $S_z\equiv 
-L_{23}$ to be the members of the Cartan subalgebra of mutually commuting 
generators. Here $N_e$, the number of electrons and $M$, the number of 
lattice sites are both taken to be even. Moreover, the Casimir operator $% 
C=\sum_{a<b} L_{ab}^2$ commutes with all generators and can be used to label 
the level of a representation. The $\pi$ operators are defined as follows  
\begin{eqnarray} 
\pi_\alpha = \sum_{\vec{p}} g(\vec{p}) 
c_{\vec{p}+\vec{Q}} (\sigma_\alpha \sigma_y) c_{-\vec{p}},
\end{eqnarray} 
where $c_{\vec{p}}$ annihilates an electron with momentum $\vec{p}$ 
(we are suppressing the spin index)
$\sigma_\alpha$ is the vector of Pauli matrices,
and $\vec{Q}=(\pi,\pi)$ is the antiferromagnetic wave vector.
If one takes $g(\vec{p})=sgn(cos p_x - cos p_y)$, the $SO(5)$ algebra closes
exactly\cite{rabello}. However, for cluster calculations it is often
more convenient to take $g(p)=cos p_x - cos p_y$, the numerical difference
between these two choices is small.
Together with the total spin raising and lowering operators $S^\pm$,
$\pi_\alpha$ and $\pi^\dagger_\alpha$  
form the root generators of $SO(5)$ and rotate different members of a 
multiplet into each other. 
 
In this paper, we are concerned with the tensorial representations of $SO(5)$.
Tensors with given symmetry types under permutation of their indices are 
classified by the Young tableaux\cite{hammermesh}. 
For the $SO(5)$ group, tensors which have 
more than two antisymmetric indices can always be mapped to tensors with 
less or equal to two antisymmetric indices by the invariant tensor $\epsilon 
^{abcde}$, the fully antisymmetric index in 5 dimensions. Therefore, all 
tensorial irreps of $SO(5)$ are characterized by two integers $(\nu ,\nu 
^{\prime })$, corresponding to the length of two rows in the Young tableaux% 
\cite{hammermesh}. 
The general $(\nu,\nu')$ series can only be constructed
from two different $SO(5)$ vectors. However, as we shall see later, 
the low lying states 
of the $t-J$ model can all be classified according to the restricted 
irreps $(\nu,0)$ generated by the superspin vector alone.
Therefore, we shall restrict ourselves to the fully symmetric tensors  
$F_{a_{1},a_{2},..,a_{\nu }}$ series $(\nu ,0)$ generated by the products of 
the $SO(5)$ vector $n_{a}$, satisfying  
\[ 
\lbrack L_{ab},n_{c}]=-i\delta _{bc}n_{a}+i\delta _{ac}n_{b}.
\] 
Here $n_a$ is the five-dimensional vector 
$(\Delta^\dagger + \Delta, \vec{S}_{\vec{Q}}, -i(\Delta^\dagger - \Delta))$,
where $\Delta = (i/2) \sum_{\vec{p}} 
c_{\vec{p}} \sigma^y c_{-\vec{p}}$ denotes the $d_{x^2-y^2}$
superconducting order parameter and $\vec{S}_{\vec{Q}}=\sum_{\vec{p}}
c^\dagger_{\vec{Q}+\vec{p}} \vec{\sigma}c_{\vec{p}}$ denotes the
antiferromagnetic Neel vector .
However, these representations are in general not 
irreducible. Since $SO(N)$ transformations preserve the norm of a vector,
the pairwise trace components of $F$ should be projected out to obtain an
irreducible tensor, i.e. $F_{a_1,a_1,..,a_\nu}=0$. Since $F$ is symmetric,
the vanishing of the first pairwise trace ensures the vanishing for all
other pairwise traces. Therefore, a pairwise traceless symmetric tensor
has  
$ 
%\begin{eqnarray}
\left(
\begin{array}{c}
N+\nu-1\\
\nu
\end{array} 
\right)
-
\left(
\begin{array}{c}
N+(\nu-2)-1\\
\nu-2
\end{array} 
\right)
%\label{Lmatrix}
%\end{eqnarray}
$
components, which gives the dimension $D_\nu$ of a level $(\nu,0)$ 
(or simply $\nu$) irreps.
For $SO(5)$ we obtain $D_\nu=\frac{1}{6}(\nu+1)(\nu+2)(2\nu+3)$, while
for $SO(3)$ $D_\nu$ reduces to the familiar degeneracy $2\nu+1$.  
The Casimir operator takes the value $\nu(\nu+3)$ for a level $\nu$ irreps.

The linear combinations of $n_a$, $\Delta^\pm=n_1\pm i n_5$,
$S^\pm_{\vec{Q}}=n_2\pm i n_3$ and $S^z_{\vec{Q}}=n_4$ are eigenvectors 
of $Q$ and $S_z$,
and can be used to label each component of a given irreps in the
two dimensional coordinate space of $Q$ and $S_z$. The diagrams of the
multiplets take the form of a diamond as
plotted in Fig. 1. Generally, a rank $\nu$
irreps contains many spin multiplets, with total spin $S=\nu$ multiplet
being the largest member. At the top of the diamond, $Q=\nu$ is a spin
singlet, at the next sublevel, $Q=\nu-1$ is a spin triplet, 
the $Q=\nu-2$ sublevel contains both a spin singlet
and a quintet. Generally, the $Q=\nu-p$ sublevel contains total spin 
$S=p,p-2,p-4,..$ multiplets. These different spin multiplets takes the
form of nested diamonds in a multiplet.
The different diamonds at level $\nu$ are stacked together to form a 
pyramid, with $\nu=0$ singlet at the apex, and the $\nu=M/2$ diamond at 
the base of the pyramid. Each member of a given irreps is a box
containing many microscopic states with the identical transformation
properties under $SO(5)$.

{\bf AF and dSC states in the SO(5) supermultiplet:} 
If we were dealing with a microscopic model with exact
$SO(5)$ symmetry\cite{rabello}, all the states at a given level $\nu$ are 
degenerate with each other at any finite systems. 
Degeneracy between different $\nu$ multiplets can only occur
in a infinite system, signaling spontaneous 
symmetry breaking  (SSB).
On a finite system, the ground state is an $SO(5)$ singlet, lying in the
$\nu=0$ box at the apex of the $SO(5)$ pyramid. The tendency towards SSB in 
the large system limit can be
recognized from the scaling of the energies of excited states with higher 
irreps. For example, in the infinite size limit, a AF state with Neel vector 
in the $xy$ plane is constructed from the linear superposition of states in 
the center column of the $SO(5)$-pyramid, while a dSC state is constructed 
from a 
linear superposition of states on the two (dSC) ridges of the $SO(5)$-pyramid. 
Because of the degeneracy within all multiplets, the AF and the dSC states 
constructed in the large system limit would have the same ground state energy.
If one applies a chemical potential term $H_\mu$, the members of a given 
multiplet with different charge quantum numbers will be linearly shifted by 
the $-2\mu Q$ term, leading to 
an equal level difference within a given multiplet. 
%In this case, level 
%crossings occur as a function of the chemical potential, and the ground state 
%would move from the apex to one of the boxes on the dSC ridges of the 
%pyramid.
%This microscopic level crossing can then be identified with the rotation of 
%the superspin from the AF to the dSC direction. 

In the microscopic Hubbard or the $t-J$ model used to describe the high-$T_c$ 
superconductors, the $SO(5)$ symmetry is not exact, and there are
different types of symmetry breaking terms. The symmetry breaking terms 
can also be classified according to irreducible tensors of the $SO(5)$ Lie
algebra. The chemical potential term mentioned above belongs to the
10-dimensional adjoint representation. The next simplest type of symmetry
breaking term preserving spin rotation and charge conservation would be
the $Q=S_z=0$ member of a 14 dimensional traceless and symmetric $(2,0)$ 
tensor, $H_g$, 
transforming like $\Delta^+ \Delta^- - 2/3 
\vec{S}_{\vec{Q}} \cdot \vec{S}_{\vec{Q}}$. This type of 
symmetry breaking has two important effects. First it can lead to 
mixing of states with $\nu$ quantum numbers differing by 2. The second
more important 
effect is the removal of the degeneracy between the members of a 
supermultiplet carrying different charge quantum numbers. However, unlike the 
chemical potential term, it preserves the
symmetry between the charge states with the same magnitude $|Q|$. This type
of symmetry breaking term can remove the degeneracy between the AF and dSC 
states when $\mu=0$, leading to a charge gap while keeping the spin
excitations at low energy. However, with an applied chemical potential, the
effects of these two types of symmetry breaking term can compensate each other
for one type of charge states, say hole states with $Q<0$, and there is a 
critical chemical
potential $\mu_c$ at which the multiplets with different charges $Q<0$
can recover their near degeneracy. As we shall see later, our overall 
numerical 
results can be consistently interpreted by these two types of explicit 
symmetry 
breaking terms. The
competition between these two types of symmetry breaking terms is analogous
to the competition between the spin anistropy and an applied uniform magnetic
field in an antiferromagnet, as illustrated in reference\cite{zhang}. 

It is important to point out that, although $H_g$ and $H_\mu$ can nearly
compensate each other on one side of the charge states, the full $SO(5)$ 
symmetry between low energy states of different signs of charge $Q$ can not be 
recovered. In the $t-J$ model for example, all states with $Q>0$ are projected 
out of the Hilbert space, and the $SO(5)$ symmetry can only be
approximately realized between the members of the supermultiplets on
the $Q<0$ half of the $SO(5)$ pyramid.   However, $Q<0$ states are the
relevant low-energy degrees of freedom in question, and the approximate
$SO(5)$ symmetry between these states is sufficient to understand the full
effect of doping. In this formalism we see the fundamental importance of the
Mott-Hubbard gap (projecting out the $Q>0$ states) on the interplay between AF
and dSC. Assuming that all the superspin multiplets are degenerate for the 
$Q<0$
states at a given $\mu_c$, one can either form a AF ordered state by the linear
superposition of the $Q=0$ members of the different $\nu$ multiplets, or 
use the same coefficients to form a pure dSC ordered state by the linear 
superposition
of the $Q=-\nu$ members of the different $\nu$ multiplets. These two states 
and the intermediate states between them\cite{schrieffer}
are degenerate at $\mu_c$. However, since 
a macroscopic number of $Q<0$ states are used to form a 
phase-coherent pure dSC state, it has a finite hole density. Therefore, Mott
insulating behavior at half-filling is compatible with the $SO(5)$ symmetry:
AF and dSC states are nearly
degenerate at $\mu_c$, but they have different density. 

{\bf Results of exact diagonalization of the t-J model:} 
Numerical demonstration of SSB requires careful study of the level spacing as 
a function of system sizes, and true SSB is generally hard to establish. 
However, it is relatively easy to recognize a nearly degenerate multiplet 
structure on a finite system. 
For example, if we have a weakly anisotropic Heisenberg model on a
lattice, the energy splitting within a multiplet would be small compared
to the splitting between multiplets. Although both level spacings may scale as
$1/M$ in the limit of large system size, the ratio of their difference
can be independent of the system size and its smallness thus can be 
recognized even with limited finite-size data. 
In the following, we study the $t-J$ model, the simplest model Hamiltonian
which incorporates the key features of the strong-correlation limit:
\[
H =  {\cal P} \; [ \; -t \sum_{\langle i,j\rangle} c_i^\dagger c_j
+ J  \sum_{\langle i,j\rangle} \vec{S}_i \cdot \vec{S}_j \;]\;{\cal P}
\]
where $\langle i,j\rangle$ denotes a summation over all nearest neighbor
pairs on a 2D square lattice,
${\cal P}$ projects onto the subspace of states with no doubly
occupied sites. The latter constraint reflects the strong correlations
in the $U/t\rightarrow \infty$ limit of the Hubbard model. 
The parameters $-t$ and $J$ are the nearest neighbor hopping and
exchange integral. We 
have numerically diagonalized the $t-J$ model on finite
lattices of $16$ and $18$ sites
(see Ref. \cite{dagotto} for pictorial representations of these 
standard systems) and studied its low lying 
eigenstates up to total spin 3 and 6 holes. In addition to their spin and 
charge quantum numbers, these states are also labeled by their total momentum 
and the point group symmetry.
In figure \ref{fig0} we show how some of the
low lying states of the $18$-site cluster $t-J$ model with $J/t=0.5$ 
fit into the 
irreps of $SO(5)$, up to the $\nu=3$ supermultiplets. The ground states within
the respective hole-number sector are labeled by an asterisk
(we note that up to now this assignment of multiplets is only a
conjecture; below we will present numerical evidence that
these groups of states indeed are $SO(5)$ multiplets).
We see that all the different
quantum numbers of the states are naturally accounted for by the 
quantum numbers of the superspin, and the levels with different charge 
$Q$ are approximately equally spaced. More precisely,
the mean level spacing within each multiplet up to
$Q=-2$ is $-2.9886$ with a standard deviation of $0.0769$.
Therefore, at a chemical potential comparable to the mean level
spacing, the superspin multiplets are nearly degenerate.
% with the difference in
%energy being $\approx -2.9$ between $Q=-2$ and $Q=0$ and
%$\approx -3.1$ between $Q=-4$ and $Q=-2$.
There are four states inside the 
``nested diamonds" which cannot yet be fully identified with our current 
methods, they are marked by the $( )$ symbol.
%%The situation is very similar for $J/t=0.25$, in that we also have
%%nearly identical energy difference (mean value: -4.2613,
%%standard deviation: 0.1074), whereas for $J/t\ge 1$
%%the situation is different
%%in that now e.g. the difference between $Q=-2$ and $Q=0$ can be written in the
%%form $a + \nu b$, with the linear coefficient $b$ being
%%$0.3$ for $J/t=1$ and $0.9$ for $J/t=2.0$. This indicates
%%that the perturbing terms for the $SO(5)$ symmetry do have a marked
%%dependence on $J/t$.

Now we wish to demonstrate that the different states inside a given multiplet
can indeed be rotated into each other by the $SO(5)$ root generators.  
In particular, we would like to show explicitly how higher spin AF
states are rotated into the dSC states.   
Let us first briefly discuss the selection rules within figure \ref{fig0}.
The $\pi$ and $\pi^\dagger$ operators act as raising and
lowering operators and obey the selection rule $\Delta \nu=0$.
In other words, they give transitions within the
`diamonds' in figure \ref{fig0}.
In the presence of an $H_g$-type of perturbation $\Delta \nu=2$ transitions
are also possible, but are expected to have smaller amplitude.
Next, as mentioned above, the $5$ operators $\Delta$,
$\Delta^\dagger$ and $\vec{S}_{\vec{Q}}$ together form an
$SO(5)$ vector. Consequently
they play a role analogous to the dipole operator
in $SO(3)$. They obey the selection rule $\Delta \nu=\pm1$,
i.e. they allow transitions between the different
diamonds. Thereby
$\Delta^\dagger$ is a spin singlet, transfers zero
momentum and has $B_1$ symmetry, whereas
$\vec{S}_{\vec{Q}}$ is a spin triplet, transfers momentum
$(\pi,\pi)$ and has $A_1$ symmetry.
$\vec{S}_{\vec{Q}}$ is
the operator relevant for neutron scattering, this experiment
thus probes $\Delta \nu=\pm1$ transitions.\\
As a `diagnostic tool' to judge
if a transition from a given state $|\Psi\rangle$
by the operator $\hat{O}$ (which can be $\pi$, $\pi^\dagger$, 
$\Delta$ or $\vec{S}_{\vec{Q}}$)
is possible, we compute the spectral function 
\begin{equation}
A_\alpha(\omega) = \Im \frac{1}{\pi} \langle \Psi|
\hat{O}^\dagger \frac{1}{\omega - (H- E_{ref}) - i0^+}
 \hat{O}| \Psi \rangle
\label{def}
\end{equation}
where $\Im$ denotes the imaginary part and
$E_{ref}$ is a suitably chosen reference energy.
For finite systems spectral functions of the type (\ref{def}) can be 
calculated exactly by means of the Lanczos 
algorithm\cite{dagotto}.
An intense and isolated
low energy peak in (say) the $\pi$-spectrum (\ref{def})
then indicates that there is a state
with $Q+2$ into which $|\Psi\rangle$ is being transformed by $\pi$.
This should hold at least as long as the 
doped state is sufficiently low in energy to be still
within the range of validity of the approximate $SO(5)$ symmetry.
Our strategy for the following,
therefore, is to work ourselves through Figure \ref{fig0}
and to check `$SO(5)$-allowed' and `$SO(5)$-forbidden' transitions by 
computing the
respective spectral functions. We can then also investigate
the influence of perturbations which could possibly
break the $SO(5)$ symmetry by studying their influence on the spectra,
to see e.g. how `$SO(5)$-forbidden' transitions are enhanced
by the perturbation.\\
We now begin to discuss the results of our `computer spectroscopy'.
Figure \ref{data1} compares the spin correlation function
at half-filling to the spectra of the $\pi^\dagger$-operator
for the lowest $2$-hole states with $d$-symmetry and
total spin $0$ and $2$. The reference energy $E_{ref}$ has been taken
as the energy of the half-filled ground state.
The spin correlation function
has a single dominant low-energy peak at an excitation
energy of $\approx 0.5J$ (marked by an arrow) which clearly should be 
associated with a magnon state (all cluster states are
exact eigenstates of $\vec{S}^2$ so there is no SSB in the
small cluster - which explains the finite excitation energy).
The spectrum of the $\pi^\dagger$-operator
for the $2$ hole, $^1B_1$ state also shows a single, high intensity
peak, which coincides with that of the spin correlation function,
i.e. the excitation energies agree within computer accuracy
($10^{-13}$). Obviously the final states are identical,
which shows that the $\pi^\dagger$-operator indeed produces the
spin resonance.
Next, the spectrum of the $\pi^\dagger$-operator
for the $2$ hole, $^5B_1$ state has a strong peak at high energy,
plus a low-energy peak
with significantly lower intensity, which again coincides with the
spin resonance. 
Here it should be noted that the transition from the
$^5B_1$ state with momentum $(0,0)$ to the $^3A_1$ state with momentum
$(\pi,\pi)$ has $\Delta \nu =2$, it is therefore an $SO(5)$ forbidden
transition, but it is allowed by spin, momentum or
point group symmetry. Inspection shows, that the intense
high energy peak in the $^5B_1$ spectrum is nothing but the
lowest $^7A_1$ state - this $\Delta \nu=0$ transition obviously is allowed by
the $SO(5)$ selection rule, see Figure \ref{fig0}. The $SO(5)$
selection rule thus is obeyed approximately, with the
ratio of the two peaks in the $^5B_1$ spectrum being a rough
measure for the degree of symmetry breaking. The pattern of the
explicit symmetry breaking is consistent with that of a second rank
$SO(5)$ tensor $H_g$. The intensity
of the peaks in the various `$\pi$-spectra' decreases
rapidly with decreasing $J/t$ - this indicates that corrections to
the $\pi$-operator become more important at smaller $J/t$.
On the other hand the additional peaks at higher energy in the
$\pi$-spectra decrease rapidly as well and always stay well-separated
in energy - restricting the Hilbert space to states
below a cut-off frequency $\approx 2J$ would therefore
give a very good eigenoperator of the Hamiltonian.\\
We proceed to the $2$-hole subspace,i.e. $Q=-1$.
Figure \ref{data2} shows the spin correlation function
for two holes, as well as various spectra of the
$\pi$ and $\pi^\dagger$ operator, $E_{ref}$ is the energy
of the two-hole ground state.
To begin with, the spin correlation function again has a dominant low
energy peak, whose excitation energy scales approximately with $J$.
The final state responsible for this peak is the lowest
$^3B_1$ state with momentum $(\pi,\pi)$.
Then, the spectrum of the $\pi$ operator calculated for
the undoped $^5A_1$, $\vec{k}=(0,0)$ state,
and the spectrum of the $\pi^\dagger$ operator
for the $^1A_1$ $\vec{k}=(0,0)$ state with $4$ holes
also do have intense low energy peaks. These peaks are
well separated from some incoherent high-energy continua, which start
above a lower bound of $\approx 2J$,
and again coincide to computer accuracy with the $^3B_1$ state
observed in the spin correlation function.
This again confirms the interpretation of the spin resonance
as a `$\pi$-excitation'.
Looking at Figure \ref{fig0} it becomes obvious that these two
transitions have $\Delta \nu=0$, i.e. they are `$SO(5)$-allowed'.
On the other hand, the $\pi$-spectrum for the undoped
{\em ground state}, $^1A_1$, has a weaker peak at the position of the
spin resonance. This $\Delta\nu=2$ transition is forbidden by the ideal
$SO(5)$ symmetry (see Figure \ref{fig0}),
indicating again a weak breaking of
the $SO(5)$ symmetry. The decrease of the `$\pi$-peaks' with decreasing
$J/t$ is quite analogous as in the case of half-filled final states.
The only exception are the peaks in the $\pi^\dagger$ spectra
(i.e. with initial states in the $4$-hole subspace), which have
practically zero weight for smaller $J/t=0.25$.\\
This also becomes
clear if we study spectra with final states in the $Q=-2$ subspace.
Figure \ref{data3} shows the spin correlation function at $4$ holes,
together with spectra of the $\pi$ operator for the
$^1B_1$ and $^5B_1$ states of two holes with momentum $(0,0)$.
>From Figure \ref{fig0}, we see that the transition
from the singlet state is forbidden, that from the quintet
is allowed. Then, looking at Figure \ref{fig0} it is apparent that
there occurs a drastic change for $J/t$ smaller than a cluster
dependent value. For $J/t \ge 0.5$ in the $18$-site cluster
($J/t \ge 1$ in the $16$-site cluster) we have the `standard
situation': the dominant
low energy `$\pi$-peak' for the $^5B_1$ initial state
is more intense than that for the $^1B_1$ state, indicating again a weak
breaking of the $SO(5)$ symmetry. Both `$\pi$-peaks'
coincide with the dominant low energy peak in the spin correlation function,
which in turn stems from the lowest $^3A_1$ state at $(\pi,\pi)$,
which confirms that the interpretation of this peak as a `$\pi$-resonance'
is valid throughout the low doping regime. On the other hand,
for smaller $J/t$ the correspondence between the
spin correlation function and the $\pi$-spectra is essentially lost,
in the case of the $16$-site cluster the spin correlation function does not
have a distinguishable low energy peak at all.
Quite obviously, we have reached the limit of applicability of the
$SO(5)$ symmetry, which seems to occur at a doping level of 
$\approx 0.25$\%, with some dependence on the ratio $J/t$.
This is roughly the same parameter range where dSC correlations vanish
on the finite size cluster.\\
Summarizing the study of the spin correlation function,
we may say that the data are in overall agreement with
an approximate $SO(5)$ symmetry, in that `$SO(5)$-allowed transitions'
usually have a larger intensity than the `forbidden' ones.
The data also show that the dominant low energy spin excitation at
$(\pi,\pi)$ always can be generated by adding or removing
two electrons from the system by means of the $\pi$-operator, which
obviously supports the conjecture of Demler and Zhang\cite{demler}
that this low energy resonance in the dynamical spin correlation function
is the hallmark of the approximate $SO(5)$ symmetry. The agreement
with the $SO(5)$ symmetry deteriorates for higher doping levels
and/or smaller $J/t$. \\
We now proceed to map some additional transitions
within the $SO(5)$ multiplets.
Figure \ref{data4} shows the spectrum of the $\pi$-operator
for the lowest triplet state with $2$ holes (this state
is the one which gives rise to the prominent peak in the
spin correlation function in Figure \ref{data2}).
The initial state thus is $^3B_1(\pi,\pi)$ with $2$ holes,
and on the basis of Figure \ref{fig0} we expect a strong
transition to the $^1A_1(0,0)$ state with $4$ holes (i.e. the
$4$ hole ground state). As the reference energy we choose
the energy of the $4$ hole ground state, and Figure \ref{data4}
then clearly shows a pronounced peak in the $\pi$-spectrum
with zero excitation energy, precisely as expected
on the basis of the $SO(5)$ symmetry. 
Figures \ref{data1}-\ref{data4}
thus demonstrate that the $4$-hole ground state can be obtained
by $2$-fold `$\pi$-rotation' from the lowest $^5A_1$ state with
momentum $(0,0)$ at half-filling. Similarly,
the $2$-hole ground state can be obtained by $\pi$-rotation
from the lowest half-filled $^3A_1$ state with momentum $(\pi,\pi)$.\\
To summarize this section, we have shown that the transitions
induced by the $\pi$ and $S(\vec{Q})$ operators can be well
understood in the framework of a weakly broken $SO(5)$ symmetry.
We have explicitly identified the multiplets with $\nu=0,1,2,3$
and shown that the $\pi$-operator gives transitions between the
members of a multiplet with different $Q$. Moreover, we have
verified explicitly that the spin correlation function
`operates in the same subspace' as the $\pi$-operator,
and that the prominent low energy peaks in the dynamical
spin correlation function corresponds to members of the
$SO(5)$ multiplets for all dopings $\leq 25$\%.

{\bf Conclusions:}
We have numerically diagonalized the low lying states
of the $t-J$ model near half-filling and found that they fit into 
irreps of $SO(5)$ symmetry group. 
At a critical value of the chemical potential, 
the superspin multiplets are nearly degenerate, and therefore higher
spin AF states at half-filling can be freely rotated into dSC states away 
from half-filling. There are clearly visible effects of $SO(5)$ symmetry
breaking, which to the lowest order can be identified with the type of 
a symmetric traceless rank two tensor. Our
overall result suggest that the low-energy 
dynamics of the $t-J$ model can
be described by a quantum $SO(5)$ nonlinear $\sigma$ model with
anisotropic couplings, and the transition from AF to dSC phase can
be identified with that of a superspin flop transition\cite{zhang}. 
It is truely remarkable that while the physical properties of AF and dSC
states are intrinsically different and they are characterized by very
different forms of order, there exists nevertheless a fundamental
$SO(5)$ symmetry that unifies them. The dichotomy between their 
apparent difference and fundamental unity is in our view a 
key which can unlock the mystery of the high $T_c$ superconductivity.  

We would like to thank Drs. J. Berlinsky, E. Demler, H. Fukuyama, S. Girvin, 
C. Henley, M. Imada, C. Kallin,
H. Kohno, R. Laughlin, S. Meixner and C. Nayak for useful discussions.
This work is supported by FORSUPRA II, BMBF (05 605 WWA 6),
ERB CHRXCT940438 and the NSF under grant numbers 
DMR-9400372 and DMR-9522415.

{\bf Appendix: Perturbations to the $t-J$ Model and Their Influence
on the Approximate $SO(5)$ Symmetry}
Recently, Baskaran and Anderson \cite{baskaran} raised some questions
concerning the effect of the diagonal hopping and nearest neighbor Coulomb
interaction on the approximate $SO(5)$ symmetry.
It is then of important to check
whether these perturbations are essentially irrelevant
or they could lead to a breakdown of the (approximate) $SO(5)$ symmetry.
Again, we resort to exact diagonalization calculations
to address this question. To begin with, we
consider the effect of an extra `Coulomb repulsion'
between holes on nearest neighbors.
More precisely, we add the term $H_V = V\sum_{\langle i,j\rangle}
n_i n_j$ to the Hamiltonian, where $n_i$ denotes the electron density
on site $i$.
Figure \ref{fig7} then compares some
spectra of the $\pi$-, $\pi^\dagger$-operator and the zero momentum 
pair-operator $\Delta$ for different values of $V$.
The left panel shows spectra with final states in the
two-hole sector, the reference energy is that of the
half-filled ground state.
With this choice the excitation energies of the dominant low-energy peaks
in both spectra increase with $V$. This is natural because the
repulsion $V$ is not operative at half-filling, but will tend to
increase the energies of hole-doped states.
The increase, however, is
significantly less than expected, being only approximately 
$0.5t$ for $V$$=$$2t$.
This can hardly come as a surprise, because we have
$\partial E_0/\partial V = 4 \langle n_i n_{i+\hat{x}} \rangle$, i.e. the
nearest neighbor density correlation function of holes. The latter
quantity is quite small in the physical range of parameters, so $V$
does not have a great impact. {\it More importantly, the difference
of the excitation energies of the $\Delta$ and $\pi$ operators is
practically independent of $V$}.
This difference of excitation energies would give the energy
required to remove a $k=(0,0)$, $d$-wave singlet pair
from the system and reinsert a $k=(\pi,\pi)$ $d$-wave triplet pair.
This is exactly what is done in a neutron-scattering
experiment - a Cooper pair from the condensate is turned into a
$\pi$-pair while scattering an incoming neutron. 
The energy difference of the peak energies in Figure \ref{fig7} thus
should correspond to the energy of the peak in
the inelastic neutron scattering cross section, and
Figure \ref{fig7} clearly shows that even rather strong
repulsion between the holes leaves this energy unchanged.
Moreover, we note that the weight of the peaks decreases only slightly
with $V$ - the decrease is also very similar for the $\Delta$ and
$\pi$ operator; this would suggest that as long as superconductivity
`survives' the influence of $V$, so does the $\pi$-resonance.
It should also be noted that the figure actually compares an
`$SO(5)$ forbidden' transition
(from the half-filled $^1A_1(0,0)$ state to the
$^3B_1(\pi,\pi)$ with two holes), and an `$SO(5)$ allowed' transition
(from the  $^1A_1(0,0)$ $4$ hole ground state to the $^3B_1(\pi,\pi)$ 
state). The ratio of intensities for both transitions
is $\approx 1/3$ and it stays so
more or less independently of $V$. This indicates that
the degree of symmetry breaking is not affected significantly
by $V$.
Next, the right-hand panel in Figure \ref{fig7}
compares the $\pi$ and $\Delta$ spectra calculated for the
ground state with two holes - it shows similar features, in particular
the difference of excitation energies is independent of $V$, and the
weights of the peaks decrease in a very similar fashion
with $V$.\\
We now consider the influence of a next-nearest neighbor hopping
integral $t'$. We choose a $t'$ between $(1,1)$-like neighbors with opposite 
sign as $t$; for noninteracting electrons this would produce
a Fermi surface similar to the LDA predictions.
Figure \ref{fig8} again compares
the spectra of the $\pi$-operator and the $\Delta$-operator
with different $t'$ and different doping levels.
The reference energy again is the ground state
energy at half-filling in the left panel, which shows spectra with
final states in the $Q=-1$ subspace;
in the right panel, which shows final states in the $Q=-2$
subspace $E_{ref}$ is the ground state energy of two holes.
The overall picture is comparable
to that seen in Figure \ref{fig7}, i.e. the difference in excitation
energies is nearly independent of $t'/t$, and in fact
even decreases with increasing $t'/t$.
In the spectra with $4$-hole final states this obviously
leads even to a kind of level-crossing, in that the
lowest `$\pi$-peak' comes down below the
lowest $\Delta$-peak for large $t'$. The intensity
of both low-energy peaks decreases in an essentially similar fashion
with increasing $|t'/t|$. One can, however, realize a kind of `crossover'
between $|t'/t|=0.1$ and $|t'/t|=0.2$, where the
spectral weight of the $\pi^\dagger$-spectra drops sharply.
The ultimate reason is a level crossing in the
$4$ hole sector from the `$SO(5)$-compatible' $^1A_1$ ground state 
to a $^1B_1$ ground state, which occurs in between these
two values of $|t'/t|$. We note that $4$ holes in $18$ sites
correspond to a hole density of $22$\%, which is nominally
far overdoped - the drop in the $\pi^\dagger$-spectra thus is not
really a reason for concern. 
Moreover, the peak in the spectrum of the $\pi$-operator 
(which is not affected by the level crossing) stays well defined,
its intensity decreasing slightly and in proportion to that of the 
`$\Delta$-peak'. All in all it is obvious that larger values of $t'$ degrade
the $SO(5)$ symmetry. On the other hand, practically all of our data show an 
intimate relationship between the $d$-wave pairing amplitude
and the $\pi$-amplitude. If the $\pi$ resonance is suppressed,
be it due to high doping, large $V$ or large $t'$, so is usually
the $d$-wave pairing. It is then only natural to conclude that as long as
the $d$-wave pairing `survives' the influence of perturbations,
so does the $\pi$-resonance.
%%%%%%%%%%%%%%%%%%%%%%%%%%%%%%%%%%%%%%%%%%%%%%%%%%%%%%%%%%%%%%%%%% 
 
%%%%%%%%%%%%%%%%%%%%%%%%%%%%%%%%%%%%%%%%%%%%%%%%%%%%%%%%%%
\begin{figure}
\caption[]{The upper diagram illustrates a general level $\nu$ irreps of
$SO(5)$. Every state can be labeled by $Q$ and $S_z$. The maximal charge is
$Q=\pm \nu$. The states labeled by a $\times$ form the shape of a diamond,
while states inside the nested diamonds are labeled by $\circ$ and $\triangle$.
Overlapping states with same $Q$ and $S_z$ are distinguished by their 
$S$ quantum numbers. The lower diagrams are for $\nu=1,2,3$ irreps of $SO(5)$.
The figure shows the energies of some low energy states for the $18$-site 
cluster
with $J/t=0.5$. The states are grouped into different
multiplets and are labeled by the spin, point group symmetry,
and total momentum. $A_1$ denotes the totally symmetric,
$B_1$ the $d_{x^2-y^2}$-like representation of the
$C_{4v}$ symmetry group. The $( )$ symbol denotes
as yet unidentified members of the respective multiplet.}
\label{fig0} 
\end{figure}
%%%%%%%%%%%%%%%%%%%%%%%%%%%%%%%%%%%%%%%%%%%%%%%%%%%%%%%%%%%%
\begin{figure}
\caption[]{
Comparison of spectral functions with undoped final states:
dynamical spin correlation function for momentum transfer
$\vec{Q}$, calculated for the half-filled $^1A_1(0,0)$
ground state;  spectrum of the
$\pi^\dagger$-operator, calculated for the $^1B_1(0,0)$
ground state in the $Q=-1$ sector; spectrum of the
$\pi^\dagger$-operator, calculated for the lowest $^5B_1(0,0)$
state with $Q=-1$. Data are shown for
different cluster sizes and values of the ratio $J/t$.}
\label{data1} 
\end{figure}
%%%%%%%%%%%%%%%%%%%%%%%%%%%%%%%%%%%%%%%%%%%%%%%%%%%%%%%%%%%%
\begin{figure}
\caption[]{
Spectral functions with final states in the $Q=-1$ subspace:
dynamical spin correlation function for momentum transfer
$\vec{Q}$, calculated for the $^1B_1(0,0)$
ground state;  spectrum of the
$\pi^\dagger$-operator, calculated for the $^1A_1(0,0)$
ground state in the $Q=-2$ sector; spectra of the
$\pi$-operator, calculated for the half-filled $^1A_1(0,0)$
ground state and the lowest half-filled $^5A_1(0,0)$ state.
}
\label{data2} 
\end{figure}
%%%%%%%%%%%%%%%%%%%%%%%%%%%%%%%%%%%%%%%%%%%%%%%%%%%%%%%%%%%%
\begin{figure}
\caption[]{
Spectral functions with final states in the $Q=-2$ subspace:
dynamical spin correlation function for momentum transfer
$\vec{Q}$, calculated for the $^1A_1(0,0)$
ground state;  spectra of the
$\pi$-operator, calculated for the $^1B_1(0,0)$
ground state and the lowest $^5B_1(0,0)$ state in the
$Q=-1$ subspace.}
\label{data3} 
\end{figure}
%%%%%%%%%%%%%%%%%%%%%%%%%%%%%%%%%%%%%%%%%%%%%%%%%%%%%%%%%%%%
\begin{figure}
\caption[]{Spectrum of the $\pi$-operator in the `spin resonance state'
$^3B_1(\pi,\pi)$ at $Q=-1$; this state is the final state
corresponding to the dominant peak in Figure \ref{data2}.}
\label{data4} 
\end{figure}
%%%%%%%%%%%%%%%%%%%%%%%%%%%%%%%%%%%%%%%%%%%%%%%%%%%%%%%%%%%%
\begin{figure}
\caption[]{Spectra of the $\pi$ and $\Delta$-operator for different
$V$. }
\label{fig7} 
\end{figure}
%%%%%%%%%%%%%%%%%%%%%%%%%%%%%%%%%%%%%%%%%%%%%%%%%%%%%%%%%%%%
\begin{figure}
\caption[]{Spectra of the $\pi$ and $\Delta$-operator for different
$|t'/t|$. $t'$ has opposite sign as $t$.}
\label{fig8} 
\end{figure}
%%%%%%%%%%%%%%%%%%%%%%%%%%%%%%%%%%%%%%%%%%%%%%%%%%%%%%%%%%%%

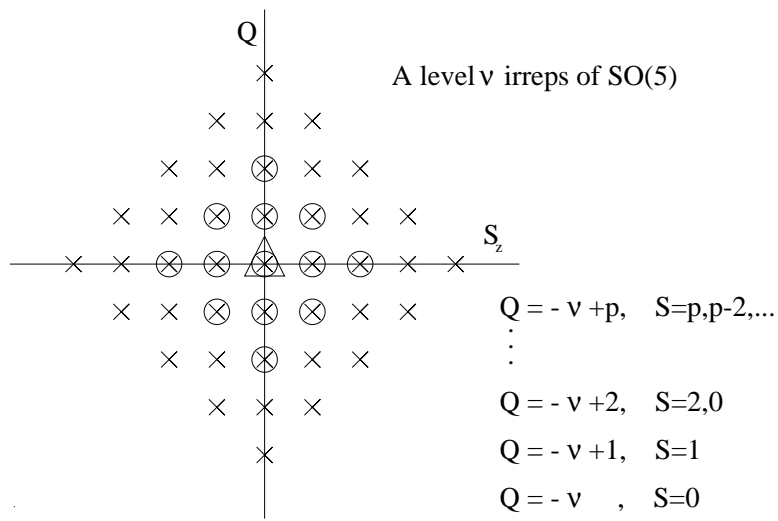
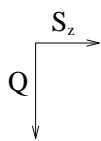
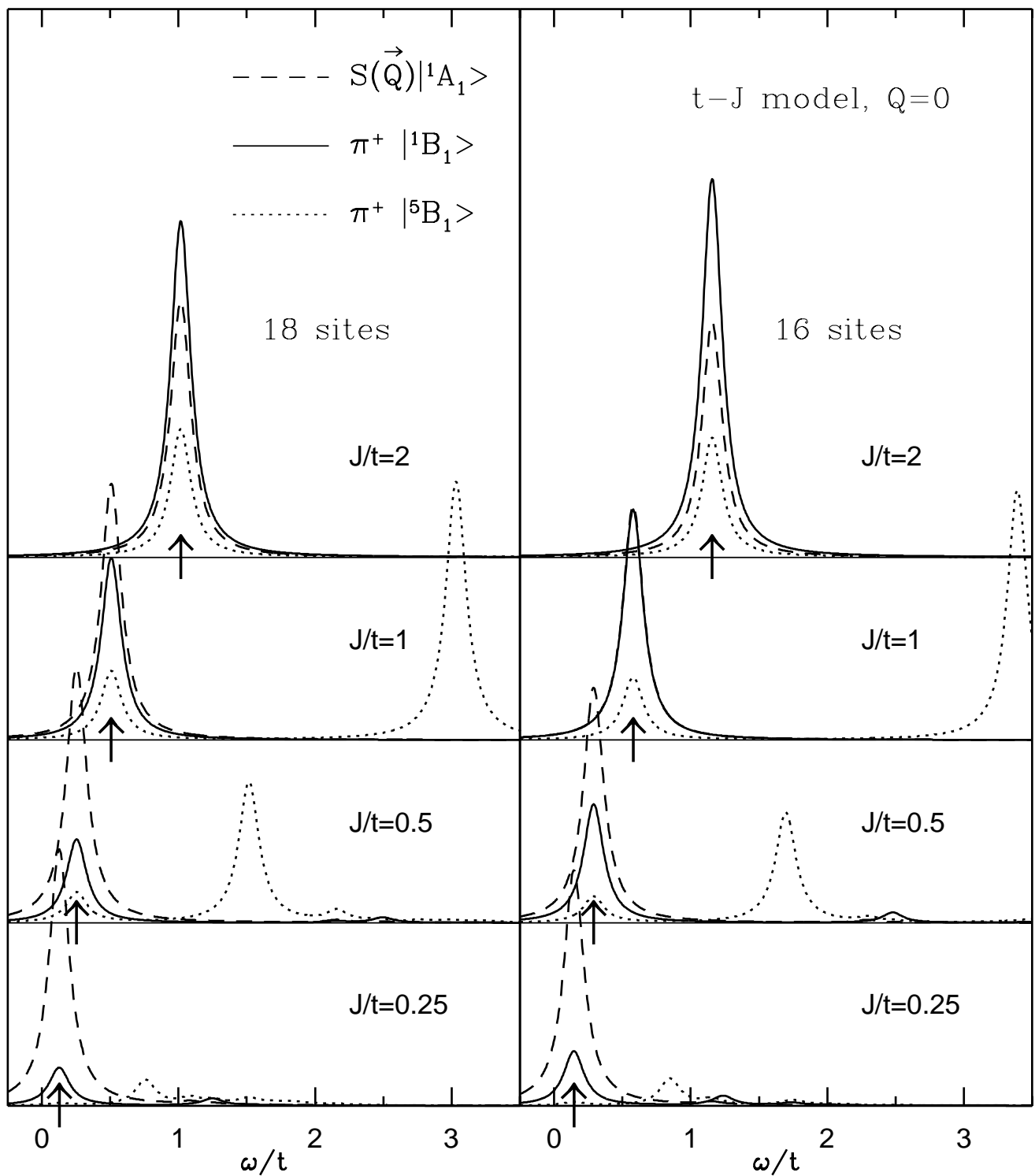
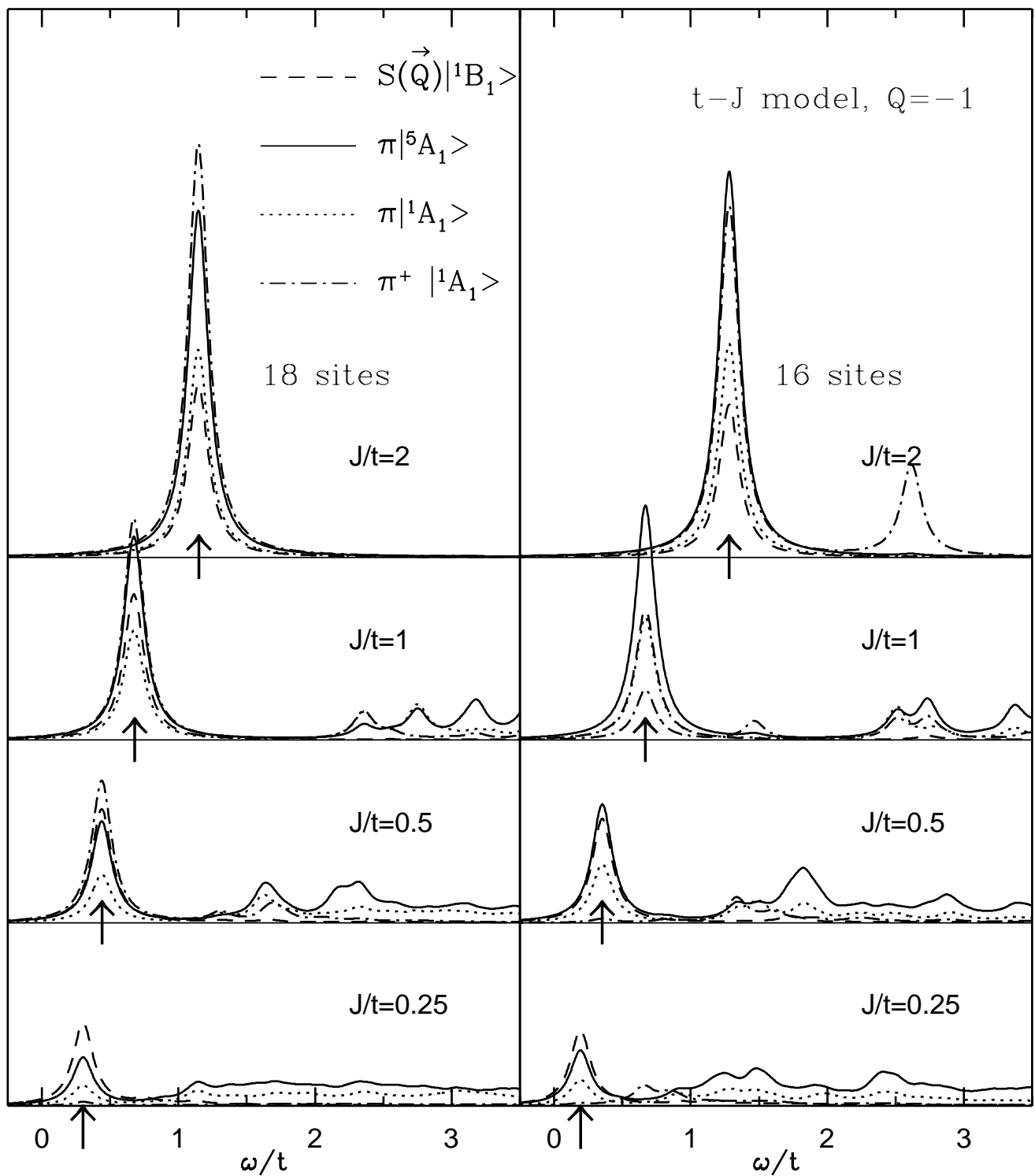
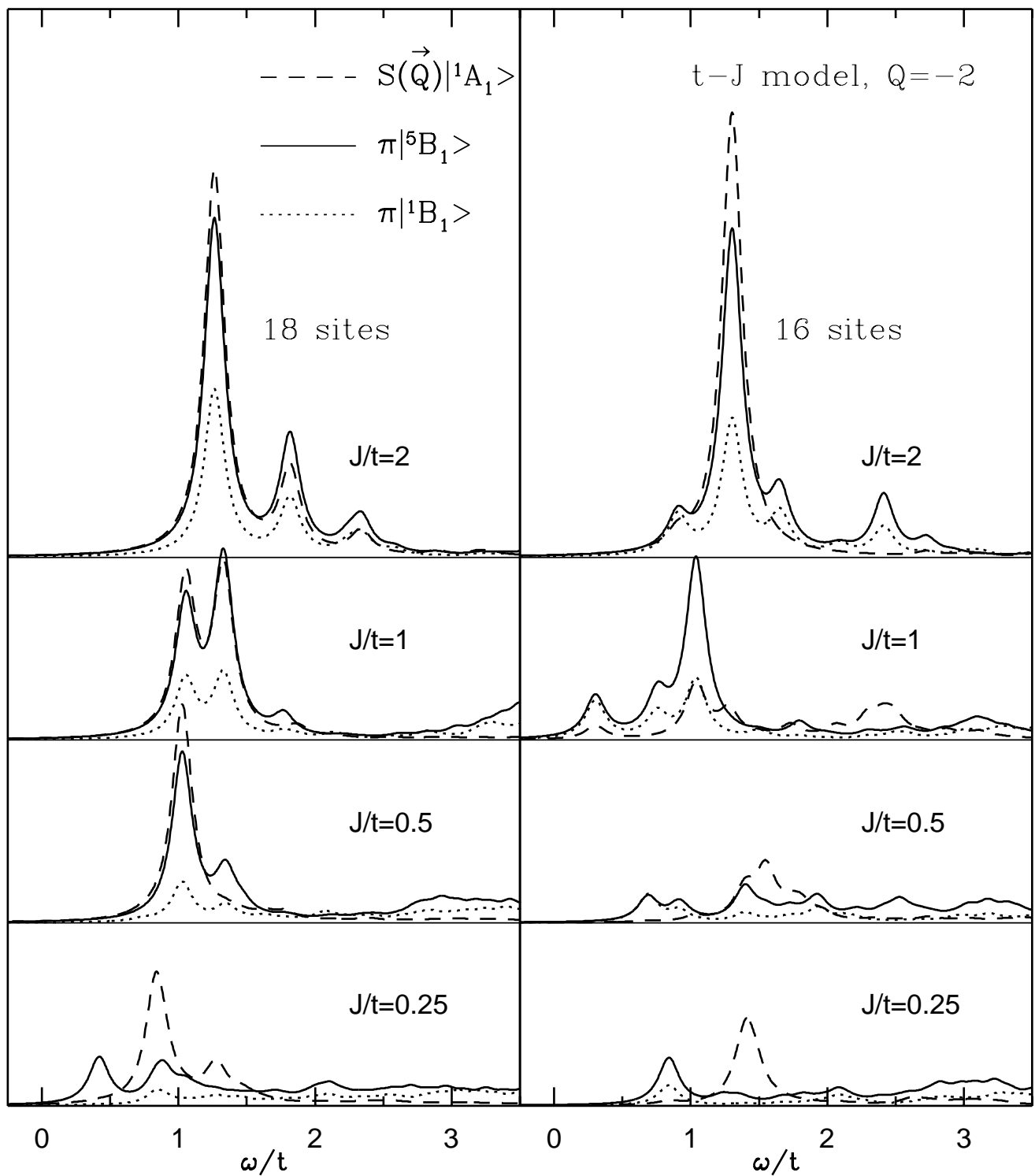
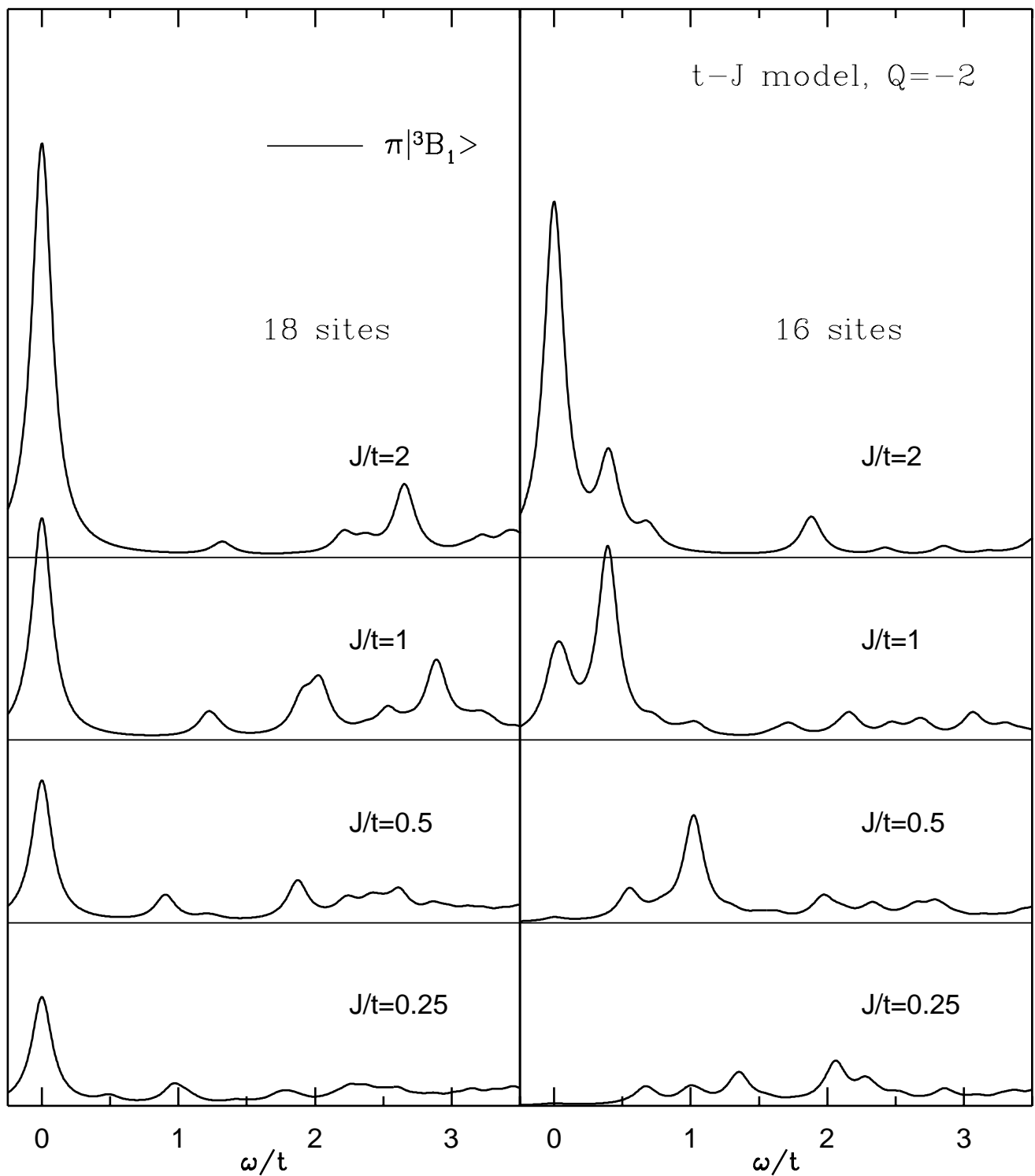
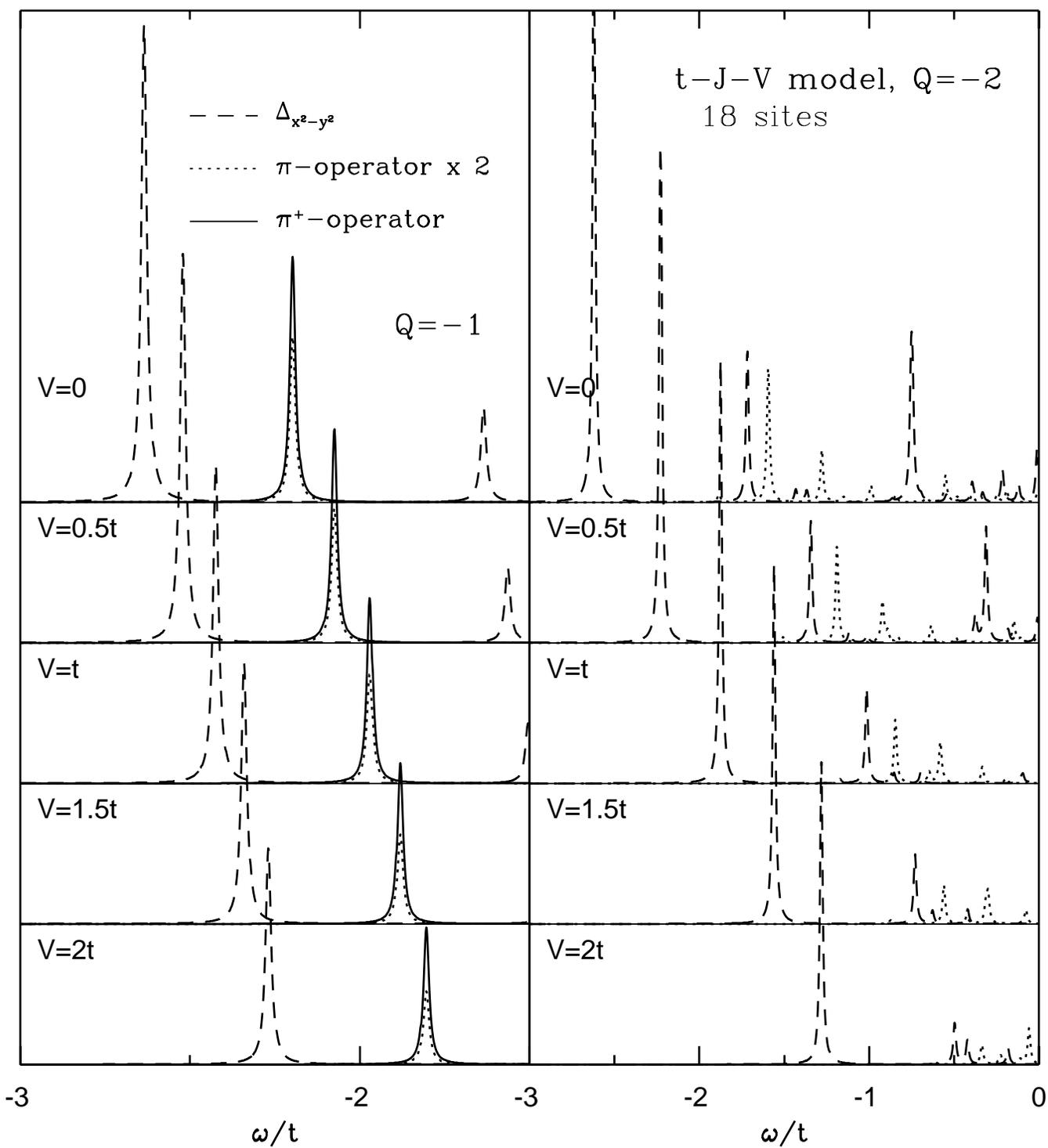
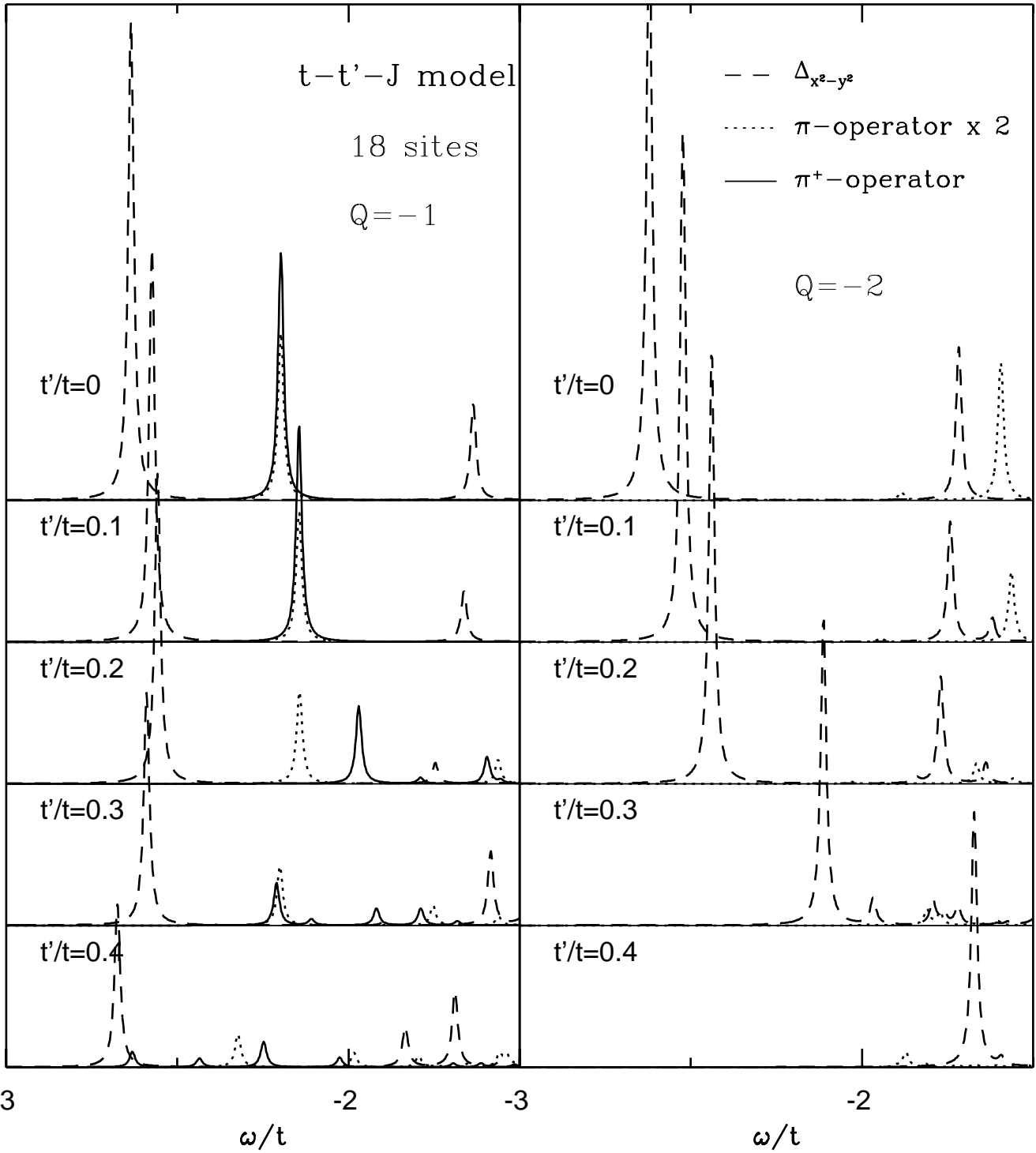
\end{document}